\documentclass[floatfix,showpacs,superscriptaddress,prl,twocolumn]{revtex4}
\usepackage{amsbsy,amssymb,amsmath,bm}
\usepackage{epsf}
\usepackage{graphicx,color}
\usepackage{amsfonts}
\usepackage{amssymb}
\usepackage{amsmath}
\def \beq {\begin{equation}}
\def \eeq {\end{equation}}
\def \bea {\begin{eqnarray}}
\def \eea {\end{eqnarray}}
\begin{document}

\title{Measurement of Magnetic Susceptibility of Diamagnetic Liquids
Exploiting the Moses Effect }
\author{D. Shulman}
\affiliation{\textit{Department of Chemical Engineering, Ariel University, Ariel 40700,
Israel}}
\affiliation{\textit{Physics Department, Ariel University, Ariel 40700, Israel}}
\author{M. Lewkowicz}
\affiliation{\textit{Physics Department, Ariel University, Ariel 40700, Israel}}
\author{E. Bormashenko}
\affiliation{\textit{Department of Chemical Engineering, Ariel University, Ariel 40700,
Israel}}
\email{edward@ariel.ac.il}
\date{\today }

\begin{abstract}
A novel comparative technique enabling measurement of the magnetic
susceptibility of diamagnetic liquids with the \textquotedblleft Moses
effect\textquotedblright\ is presented. The technique is based on the
experimental establishment of the deformation of the liquid/vapor interface
by a steady magnetic field. The deformation of the liquid surface by a
modest magnetic field $(B\sim 0.60T)$ is measured with an optical technique.
The surface tension of the liquid is taken into account. The magnetic
susceptibilities of the investigated liquids (Ethanol and Glycerol) were
calculated from the maximal slope of the liquid/air interface. The suggested
approach yields an accuracy on the order of about $0.4-0.6\%$ and hence is
superior to previous methods which used the \textquotedblleft Moses
effect\textquotedblright\ for the same purpose.
\end{abstract}

\maketitle

\section{Introduction}

Measurement of magnetic susceptibility of liquids in a non-trivial
experimental task \cite{1,2,3,4,5,6,7}. The standard methods used to measure
the magnetic susceptibility of feebly magnetic objects are based on the
measurement of the force acting on the object when placed in a non-uniform
magnetic field. In all these methods it is essential that the body can be
displaced. The force acting on the body is small, usually only a few dynes,
and delicate methods have to be used for achieving accurate measurements, as
in the common Curie-Faraday method where the studied object is placed on a
sensitive torsion balance \cite{2}, \cite{5}. Several methods were used for
the measurement of the magnetic susceptibility such as Faraday's scale,
Guoy's scale or inductive method exploiting the SQUID magnetometer \cite{2}.
A nuclear magnetic resonance (NMR) method was effectively used for measuring
the magnetic susceptibility of bar-shaped samples that have an arbitrary
cross-section and do not produce the MR signal \cite{4}, \cite{6} . Magnetic
susceptibility of a broad variety of materials was studied systematically
with a magnetic resonance recently in ref. \cite{7}.

Measurement of magnetic susceptibility of liquids poses additional
experimental problems. These problems were surmounted by Broersma, who
applied the torsion balance and an inductance apparatus for the precise
measurement of magnetic susceptibility of diamagnetic organic liquids \cite%
{1}. In parallel, magnetic susceptibility of diamagnetic organic liquids was
studied with NMR in ref. \cite{7}. Recently the method of the measurement of
diamagnetic liquids based on the Moses effect was introduced \cite{3}. Moses
effect is a phenomenon of deformation of the surface of a diamagnetic liquid
by a magnetic field \cite{8,9,10,11,12,13,14}.

Moses effect is a relatively weak phenomenon: the application of magnetic
fields on the order of magnitude of $0.1-1T$ results in the formation of a
near-surface \textquotedblleft well\textquotedblright\ with a depth of
dozens of micrometers \cite{10}. However, it is sufficient for the
measurement of the magnetic susceptibility of the studied diamagnetic
liquids, as shown in ref. \cite{3}, in which the deformation of the liquid
surface by a magnetic field was determined by a laser bounce. The analysis
of the shape of the near-surface \textquotedblleft well\textquotedblright\
produced by the magnetic field yielded the value of the magnetic
susceptibility. We demonstrate, how the accuracy of the Moses Effect method,
introduced in ref. \cite{3}, may be essentially improved, under considering
interfacial effects, namely the energy of the deformed liquid/vapor
interface.

\section{Theoretical basis of the suggested experimental method}

The original theory of the Moses effect \cite{11} implies that the shape of
the \textquotedblleft dip\textquotedblright\ formed by the permanent magnet
emerges as a minimum of the sum of the gravitational and magnetic energies.
Assuming a cylindrically symmetrical magnetic field $\vec{B}\left(
r,h\right) $ the following expression is predicted \cite{3,10,15} for the
shape of the well $z(r)$, schematically shown in Fig. (1):

\begin{equation}
z\left( r,h\right) =\frac{\chi B^{2}\left( r,h\right) }{2\mu _{0}\rho g}.
\label{wrong z}
\end{equation}

$\chi $ and $\rho $ are the magnetic susceptibility and the density of the
liquid, respectively, and $h$ is the separation between the magnet and
non-deformed liquid/vapor interface.

Eq.(\ref{wrong z})neglects the surface tension of the liquid/vapor
interface. Thus, the magnetic susceptibility according to Eq.(\ref{wrong z})
inevitably includes a systematic error. We introduce a model that includes
the surface tension, so that a more accurate prediction of the magnetic
susceptibility of diamagnetic liquids becomes possible.

The Young-Laplace equation defining the shape of the liquid-air interface
deformed by a time-independent magnetic field, which takes into account
surface tension $\gamma $ is derived in \cite{10}:

\begin{equation}
\frac{\partial ^{2}z}{\partial r^{2}}+\frac{1}{r}\frac{\partial z}{\partial r%
}-\frac{\rho g}{\gamma }z=\frac{\chi B^{2}\left( r,h\right) }{2\mu
_{0}\gamma }  \label{z equation}
\end{equation}

\bigskip with its solution 
\begin{widetext}%

\begin{equation}
z\left( r,h\right) =-\left[ \int_{0}^{r}\frac{\chi B^{2}\left( r,h\right) }{%
2\mu _{0}\gamma }I_{0}\left( \lambda _{c}^{-1}r\right) rdr\right]
K_{0}\left( \lambda _{c}^{-1}r\right) -\left[ \int_{r}^{\infty }\frac{\chi
B^{2}\left( r,h\right) }{2\mu _{0}\gamma }K_{0}\left( \lambda
_{c}^{-1}r\right) rdr\right] I_{0}\left( \lambda _{c}^{-1}r\right) .
\label{z}
\end{equation}

\end{widetext}%
Here $I_{\alpha }\left( x\right) $ and $K_{\alpha }\left( x\right) $ are the
modified Bessel functions of the first and the second kind, respectively.

The interplay between the gravity and the surface tension is quantified by
the capillary length, denoted $\lambda _{c}$ (see \cite{17,18,19}), defined
as:

\begin{equation}
\lambda _{c}=\sqrt{\frac{\gamma }{\rho g}}.  \label{lc}
\end{equation}

It is noteworthy that the value of the capillary length is of the order of a
few millimeters for a majority of liquids, in particular for water $\lambda
_{c,H_{2}O}=2.71mm$ (see \cite{10},\cite{16,17,18}). The physical properties
of the liquids used in our investigation are supplied in Table 1. The values
of the capillary lengths calculated for the studied liquids are confined in
the range of $\lambda _{c}\cong 1.7-7.2mm$ (see Table 1). The typical scale
at which the shape of the near-surface well is constituted is on the order
of magnitude of $\sim 10mm$, as recognized from Figure 2; thus the effects
due to the surface tension are well expected to be essential. Measuring the
surface well profile $z(r,h)$ and obtaining the magnetic susceptibility from
Eq.( \ref{wrong z}) (which neglects the surface tension), as reported in 
\cite{3}, inevitably generates a relatively high error of $\chi $.

We suggest an experimental procedure that enables a more accurate estimate
of the magnetic susceptibility of diamagnetic liquids, due to two essential
improvements, namely:

1)\qquad Considering the effects of the surface tension and the use of the
solution of the Young-Laplace equation which yields a more faithful shape of
the liquid surface deformed by the magnetic field;

2)\qquad Use of the maximal slope of the near-surface well for the
calculation of the magnetic susceptibility of the addressed diamagnetic
liquids.

The measurement of the maximal slope angle $\theta _{m}$ of the liquid/vapor
interface enables an accurate establishment of the magnetic susceptibility.
For small curvature $dz/dr\ll 1$, so that $dz/dr\cong \theta $, see Fig. 1.
The derivative of Eq.(\ref{z}):%
\begin{widetext}%

\begin{equation}
\theta =\frac{dz}{dr}=\left[ \int_{0}^{r}\frac{\chi B^{2}\left( r,h\right) }{%
2\mu _{0}\gamma }I_{0}\left( \lambda _{c}^{-1}r\right) rdr\right] \lambda
_{c}^{-1}K_{1}\left( \lambda _{c}^{-1}r\right) -\left[ \int_{r}^{\infty }%
\frac{\chi B^{2}\left( r,h\right) }{2\mu _{0}\gamma }K_{0}\left( \lambda
_{c}^{-1}r\right) rdr\right] \lambda _{c}^{-1}I_{1}\left( \lambda
_{c}^{-1}r\right)  \label{theta}
\end{equation}

Denoting the distance from the axis of the magnet to the point of maximal
slope as $r_{m},$ and inverting Eq.(\ref{theta}), yields an expression for
the magnetic susceptibility:

\begin{equation}
\chi =\theta _{m}\cdot \left\{ \left[ \int_{0}^{r_{m}}\frac{B^{2}\left(
r,h\right) }{2\mu _{0}\gamma }I_{0}\left( \lambda _{c}^{-1}r\right) rdr%
\right] \lambda _{c}^{-1}K_{1}\left( \lambda _{c}^{-1}r_{m}\right) -\left[
\int_{r_{m}}^{\infty }\frac{B^{2}\left( r,h\right) }{2\mu _{0}\gamma }%
K_{0}\left( \lambda _{c}^{-1}r\right) rdr\right] \lambda
_{c}^{-1}I_{1}\left( \lambda _{c}^{-1}r_{m}\right) \right\} ^{-1}
\label{chi}
\end{equation}

Assuming that the space dependence of the magnetic field $\vec{B}(r,h)$ is
known, one could find $r_{m}$ from the second derivative of $z\left(
r,h\right) $ which vanishes at $\theta _{m}:$

\begin{eqnarray}
\frac{d^{2}z}{dr^{2}} &=&\frac{\chi B^{2}\left( r,h\right) }{2\mu _{0}\gamma 
}-\left[ \int_{0}^{r}\frac{\chi B^{2}\left( r,h\right) }{2\mu _{0}\gamma }%
I_{0}\left( \lambda _{c}^{-1}r\right) rdr\right] \lambda _{c}^{-2}\left(
K_{0}\left( \lambda _{c}^{-1}r\right) +\frac{1}{\lambda _{c}^{-1}r}%
K_{1}\left( \lambda _{c}^{-1}r\right) \right)  \label{z''} \\
&&-\left[ \int_{r}^{\infty }\frac{\chi B^{2}\left( r,h\right) }{2\mu
_{0}\gamma }K_{0}\left( \lambda _{c}^{-1}r\right) rdr\right] \lambda
_{c}^{-2}\left( I_{0}\left( \lambda _{c}^{-1}r\right) -\frac{1}{\lambda
_{c}^{-1}r}I_{1}\left( \lambda _{c}^{-1}r\right) \right)  \notag
\end{eqnarray}

and hence

\begin{eqnarray}
B^{2}\left( r_{m},h\right) &=&\left[ \int_{0}^{r_{m}}B^{2}\left( r,h\right)
I_{0}\left( \lambda _{c}^{-1}r\right) rdr\right] \lambda _{c}^{-2}\left(
K_{0}\left( \lambda _{c}^{-1}r_{m}\right) +\frac{1}{\lambda _{c}^{-1}r_{m}}%
K_{1}\left( \lambda _{c}^{-1}r_{m}\right) \right)  \label{rmax from B} \\
&&-\left[ \int_{r_{m}}^{\infty }B^{2}\left( r,h\right) K_{0}\left( \lambda
_{c}^{-1}r\right) rdr\right] \lambda _{c}^{-2}\left( I_{0}\left( \lambda
_{c}^{-1}r_{m}\right) -\frac{1}{\lambda _{c}^{-1}r_{m}}I_{1}\left( \lambda
_{c}^{-1}r_{m}\right) \right)  \notag
\end{eqnarray}

\end{widetext}%
The only unknown variable $r_{m}$ in Eq.(\ref{rmax from B}) can be found by
numerical methods \cite{19} \textit{if} the space dependence of the magnetic
field $\vec{B}(r,h)$ is known. Albeit, an accurate measurement of this space
dependence of the magnetic field is challenging and poses essential
experimental problems. The inaccuracy in the measurement of the magnetic
field introduces the main experimental error in the assessment of the
magnetic susceptibility of diamagnetic liquids. We can avoid this problem
using the following experimental comparative procedure: water (a diamagnetic
liquid with a well-known magnetic susceptibility $\chi _{H_{2}O}=-9.035\cdot
10^{-6}$ \cite{24}) is used as \textquotedblleft calibration
liquid\textquotedblright . The shape of the well produced in water by the
magnet and the value of $r_{m}$ c ean be stablished as described in detail
in the Experimental Section. Thus, the space dependence of the magnetic
field can be extracted from the shape of the near-surface well produced in
the \textquotedblleft calibration liquid\textquotedblright\ (water) as
described in detail in Appendix A. This magnetic field $\vec{B}(r,h)$ will
be used for the measurement of the magnetic susceptibility of other
diamagnetic liquids, as glycerol and ethanol in this study. The value of the
magnetic susceptibility is then calculated from Eq.(\ref{chi}).

\bigskip

\section{Materials and Method}

The suggested experimental technique for the measurement of the magnetic
susceptibility of diamagnetic liquids is based on the profile of the
near-surface well forced in the liquid by a permanent magnetic field. The
main elements of the experimental unit, shown in Figs.(3) and (4) are:

1)\qquad the sample liquid placed into a Petri dish $\varnothing $ $90mm$

2)\qquad a $633nm\ $He-Ne laser (4mW 1107/P JDS Uniphase Corporation)

3)\qquad a stack of neodymium permanent magnets (Magsy)

4)\qquad XYZ gantry, in-house assembled from CCM Automation Technology
components

5)\qquad digital camera, (8.0 megapixel digital bridge camera Sony
Cyber-shot DSC-F828)

6)\qquad GM2 Gauss meter (AlphaLab Inc, USA), accuracy 1\%

The Petri dish with the liquid sample filled to a depth of $1\pm 0.2cm$ was
set on top of a frame constructed from non-ferromagnetic material (see
Figs.(3) and (4)). A cylindrical bar magnet of neodymium $(NdFeB)$ with the
diameter $22.8mm$ and length $30mm$ was used. The magnetic field strength at
the base center was $\ B_{z}=0.60\pm 0.01T$.

The magnetic field was modelled as an ideal solenoid \cite{20,21,22,23}. The
measured magnetic field showed good similarity with the solenoid field.
Fig.(7) shows a comparison of both fields at a distance of $h=0.5mm$ from
the magnet's base. The location of the magnet was fixed with the XYZ
actuator with an accuracy of $10\mu m$. In the experiments an Arduino
controller for the step motors was used. Fig.(3) shows the experimental
set-up to measure the exact separation of the magnet from the liquid surface 
$h$. The minimum step of the XYZ actuator was $10\mu m.$

The position of the laser beam on the screen was registered with the digital
camera. The experiments were carried out under ambient conditions ($%
P=1atm;T=25\pm 1^{0}C$). A photograph of the experimental unit is shown in
Fig.(4).

The investigated liquids, see Table 1, were exposed to the permanent
magnetic field; the surface of the liquid was illuminated with the laser
beam as shown in Fig. (5).

The deformation profile of the liquid sample is measured with the Ne-He
laser. Its beam diameter is $0.48mm$ and the beam divergence $1.7mrad$. The
angle of incidence $\Theta $ on the surface of the liquid is about $5^{0}$.
The inclination of the curved surface was extracted from the angle of
reflection of the laser beam.

When the fluid is deformed by a magnetic field by an angle $\Delta \theta $,
the reflection angle shifts by $2\Delta \theta $, which causes a
corresponding shift in height $\Delta z$ on the screen, as shown in Fig.(5).
Hence we have the following relationship for the change in the angle of
reflection:

\begin{equation}
2\Delta \theta =\arctan \left( \frac{y+\Delta y}{r}\right) -\Theta
\label{change}
\end{equation}

The laser spot was captured on the screen with the digital camera while
displacing the magnet horizontally. Further video processing enabled the
determination of the maximal surface slope $\theta _{m}$ which in turn
yields $r_{m}$ in Eq.(\ref{rmax from B}). Substituting $\theta _{m}$ and $%
r_{m}$ into Eq.(\ref{chi}) allows the calculation of the magnetic
susceptibility of the studied liquid $\chi $.

The measurement was performed as follows: the magnet was fixed at the XYZ
precision stage, which displaced the magnet, as shown in Figs.(5) and (6).
The laser beam scanned the curved liquid/vapor interface. The XYZ stage
lifted the magnet vertically with the step of $0.1mm$ and the scanning was
repeated. The maximal slope of the liquid/air interface was established at
each height. An arithmetic mean of the magnetic susceptibility was accepted
as a statistically reliable value.

The materials used in the experiments:

1.\qquad Distilled water, prepared by using a Synergy ultraviolet water
purification system (Millipore Sas, France)

2.\qquad Ethanol (Bio-Lab Ltd., Israel)

3.\qquad Glycerol (J.T. Baker Chemicals, Holland)

4.\qquad Calcium chloride (Merck KGaA, Germany)

5.\qquad Sodium chloride (Sigma-Aldrich, Germany)%
\begin{widetext}%

\subparagraph{Table 1}

\subparagraph{\ }

\begin{equation*}
\begin{tabular}{llll}
Liquid & Density $\rho $\ $\left[ \frac{g}{cm^{3}}\right] $ & Surface
Tension $\gamma \ \left[ \frac{mJ}{m^{2}}\right] $ & Capillary length $%
\lambda _{c}\ \left[ mm\right] $ \\ 
Water & $0.998$ & $72.8$ & $2.71$ \\ 
Glycerol & \thinspace $0.126$ & $64.0$ & $7.20$ \\ 
Ethanol & $0.789$ & $22.1$ & $1.69$%
\end{tabular}%
\end{equation*}%
The liquids' parameters, at room temperature.%
\end{widetext}%

\bigskip

\section{RESULTS AND DISCUSSION}

The suggested method does not measure the magnetic field directly, which
inevitable would give rise to relatively large errors, resulting in large
errors in the measurement of magnetic susceptibility. Rather, a comparative
method of measurement was exploited: the measurement was calibrated with
water at standard conditions, which has a well-known magnetic susceptibility 
$\chi $.

The experimentally established values of the volume\ magnetic susceptibility
(in SI units) for Ethanol and Glycerol were $\chi _{ethanol}=-7.23\cdot
10^{-6}\pm 0.03\cdot 10^{-6}$ and $\chi _{glycerol}=-9.83\cdot 10^{-6}\pm
0.06\cdot 10^{-6}$, correspondingly. The deviation of the reported results
from the literature data \cite{24} was $0.6\%$ for ethanol and $0.39\%\ $for
glycerol. The comparison of the experimentally obtained and literature data
is shown in Fig.(6).

For the measurement of the magnetic susceptibility of salt solutions we
choose $NaCl\ $and $CaCl_{2}$. The experimental values for these compounds
are summarized in Table 2. Calculation of the susceptibilities of dissolved
salts has usually been based on the assumptions that the susceptibility of
the solution varies linearly with concentration and that the susceptibility
of the solvent is not affected by the presence of dissolved salts \cite{25}.
The validity of these assumptions can be verified by the experimental data,
see Fig.(6). The assumption of a linear dependence of the susceptibility of
solutions with concentration seems to be justified for many simple salts by
the results obtained. In most cases the susceptibility of the salt in the
dissolved state is greater than it in the crystalline state \cite{25}.

\subparagraph{Table 2. Volume susceptibility (in SI units)}

\begin{tabular}{llll}
& crystalline state$^{\left( 1\right) }$ & aqueous solution$^{\left(
1\right) }$ & aqueous solution$^{\left( 2\right) }$ \\ 
& $\chi _{crystal}\ \ $ & $\chi _{dissolved}^{\left( 1\right) }$ & $\chi
_{dissolved}^{\left( 2\right) }$ \\ 
$NaCl$ & $-1.40\cdot 10^{-5}$ & $-1.41\cdot 10^{-5}$ & $-1.40\cdot 10^{-5}$
\\ 
$CaCl_{2}$ & $-1.33\cdot 10^{-5}$ & $-1.36\cdot 10^{-5}$ & $-1.37\cdot
10^{-5}$%
\end{tabular}

$\left( 1\right) \ $\ from ref. \cite{25} \ \ \ \ 

$\left( 2\right) \ $\ our experimantal value

In order to demonstrate the improved accuracy of the suggested experimental
technique the effect of the surface tension was analyzed in Appendix B. The
effect is substantial for radii of the magnet smaller than the capillary
length, Eq.(\ref{lc}), for larger radii it becomes negligible, though.
Albeit, the use of \textquotedblleft giant magnets\textquotedblright\ would
render the method cumbersome.

\section{Conclusion}

The \textquotedblleft Moses Effect\textquotedblright\ enables an accurate
experimental measurement of the magnetic susceptibility of diamagnetic
liquids when interfacial phenomena effects are considered. Magnetic
susceptibility of diamagnetic liquids was already calculated from the shape
of the well created by a magnetic field in the studied liquid \cite{3}. The
method reported in ref. \cite{3} neglects the effect of the surface tension.
We improve the method suggested in ref. \cite{3} as follows: i) the effects
due to the surface tension are considered; ii) a comparative method is
suggested using water a calibration liquid; iii) the magnetic susceptibility
of the investigated liquids are found from the maximal slope of the
liquid/air interface.

\section{Appendix A.}

In order to determine the magnetic susceptibility with (\ref{chi}) one needs
the explicit spatial distribution of the magnetic field produced by the
permanent magnet, $\vec{B}(r,h)$. This distribution was approximated by the
ideal solenoid model described by the following equations \cite{20}%
\begin{widetext}%

\begin{equation*}
B_{r}\left( r,h\right) =B_{0}\int_{0}^{\pi /2}d\psi \left( \cos ^{2}\psi
-\sin ^{2}\psi \right) \left\{ \frac{\alpha _{+}}{\sqrt{\cos ^{2}\psi
+k_{+}^{2}\sin ^{2}\psi }}-\frac{\alpha _{-}}{\sqrt{\cos ^{2}\psi
+k_{-}^{2}\sin ^{2}\psi }}\right\}
\end{equation*}

\begin{equation*}
B_{z}\left( r,h\right) =\frac{B_{0}a}{r+a}\int_{0}^{\pi /2}d\psi \left( 
\frac{\cos ^{2}\psi +\tau \sin ^{2}\psi }{\cos ^{2}\psi +\tau ^{2}\sin
^{2}\psi }\right) \left\{ \frac{\beta _{+}}{\sqrt{\cos ^{2}\psi
+k_{+}^{2}\sin ^{2}\psi }}-\frac{\beta _{-}}{\sqrt{\cos ^{2}\psi
+k_{-}^{2}\sin ^{2}\psi }}\right\}
\end{equation*}

\begin{equation*}
\alpha _{\pm }=\frac{a}{\sqrt{h_{\pm }^{2}+\left( r+a\right) ^{2}}},\ \ \ \
\ \ \ \ \ \beta _{\pm }=\frac{h_{\pm }}{\sqrt{h_{\pm }^{2}+\left( r+a\right)
^{2}}}\ \ \ \ \ \ 
\end{equation*}

\begin{equation*}
h_{+}=h,\ \ \ h_{-}=h-2b,\ \ \ \ \tau =\frac{a-r}{a+r}
\end{equation*}

\begin{equation*}
k_{\pm }=\sqrt{\frac{h_{\pm }^{2}+\left( a-r\right) ^{2}}{h_{\pm
}^{2}+\left( a-r\right) ^{2}}}
\end{equation*}

\end{widetext}%
where $a$ is the radius and $2b$ is the length of the solenoid, equal to the
radius and the length of our permanent magnet, and $B_{0}$ is a constant
with units \textit{tesla}. In order to calculate the magnetic susceptibility 
$\chi $ , Eq.(\ref{chi}), one needs to know the value of$\ B_{0}$, albeit an
accurate measurement of $B_{0}$ is accompanied with considerable
experimental challenges.

To determine the numerical value of $B_{0}$ we use water as a calibration
liquid, its physical properties (namely: density $\rho $, surface tension $%
\gamma $ and magnetic susceptibility $\chi $) being well-known \cite{24}.
Measuring the maximal slope $\theta _{m}$ for water and inverting Eq.(\ref%
{theta}) results in the value $B_{0}=1.2T$. To improve the accuracy of the
measurement the experiments were repeated at different separations $h$. The
comparison of the experimentally established radial dependence of the
magnetic field with that calculated with the above equations is depicted in
Fig.(7).

\bigskip

\section{Appendix B:}

\subparagraph{Influence of the surface tension on the accuracy of the
measurement}

Various groups determined the magnetic susceptibility of diamagnetic liquids
with the Moses effect \cite{3}, where the magnetic susceptibility was
derived from the well's profile created by the magnetic field. The
calculation based on the Eq. (\ref{wrong z}) neglects the effect of the
liquid's surface tension. Here we estimate the error arising from ignoring
the interfacial effects, i.e., we compare the values of the magnetic
susceptibilities derived from Eq. (\ref{wrong z}) and Eq. (\ref{chi}).

To compare the two methods, we introduce a dimensionless number $\eta =\chi
_{1}/\chi _{2}$, the ratio between the values of the magnetic susceptibility
obtained under considering surface tension, denoted $\chi _{1}$, and
neglecting surface tension, denoted $\chi _{2}$.

An interplay between the effects due to gravity and surface tension is
quantified by the capillary length $\lambda _{c}=\sqrt{\frac{\gamma }{\rho g}%
}$ introduced in Section 2. Let us define the dimensionless number $\xi $
relating the radius of the magnet for the deformation of the liquid/vapor
interface to the capillary length defined as follows: $\xi =\left(
R_{magnet}/\lambda _{c}\right) ^{2}$. In Fig. (8) the dependence of the
accuracy of our measurement $\eta $ on the parameter $\xi $ is shown. As $%
\xi $ increases, the effect of the surface tension decreases, and the
interfacial effects become negligible.

Indeed, the interplay between the effects due to gravity and surface tension
for the addressed experimental situation may be quantified by the
dimensionless number $\zeta $ defined as follows:

\begin{equation*}
\zeta =\frac{\gamma 2\pi r}{\rho \pi r^{2}Hg}=\frac{\gamma 2}{\rho rHg}
\end{equation*}

where $r$ and $H$ are the radius and depth of the near-surface well,
correspondingly. The radius increases obviously with $R_{magnet}$ . The
parameter $\zeta $ scales as $1/r$ for the fixed depth of the well. Thus,
with the increase of the radius of the magnet the effects due to surface
tension disappear.

We conclude that the effects due to the surface tension become negligible
when the radii of the magnets are much larger than the capillary length.

\pagebreak 

\begin{figure*}[btp]
\centering
\includegraphics[scale=1]{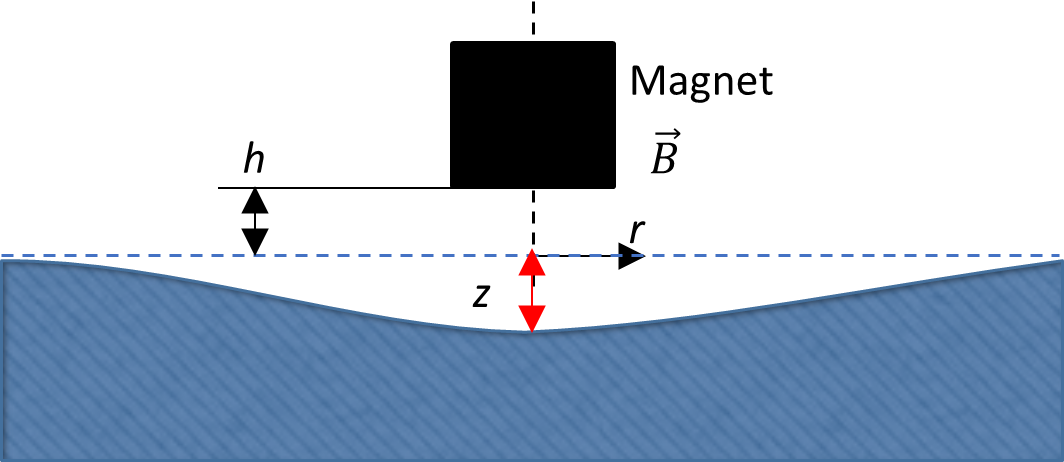}
\caption{Shape of the near-surface well arising from the deformation of the liquid/vapor interface by the permanent magnetic field  $\vec{B}$ is depicted.}
\label{fig:1}

\includegraphics[scale=1]{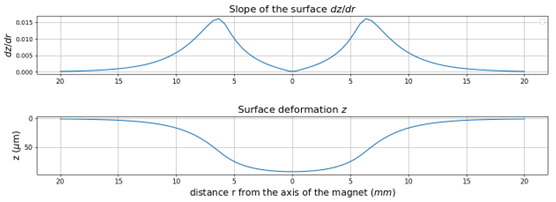}
\caption{The water curvature (lower inset) and the slope of the curvature (upper inset) for the magnet with radius $a= 3\ mm$, placed at the distance of $h= 0.5\ mm$ from the liquid/vapor interface, as a function of the distance from the magnet axis, as calculated from the Eqs. (3) and (4) respectively are depicted.}
\label{fig:2}

\includegraphics[scale=1]{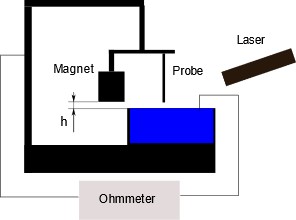}
\caption{The scheme of the experimental unit is shown.}
\label{fig:3}
\end{figure*}
\begin{figure*}
\includegraphics[scale=1]{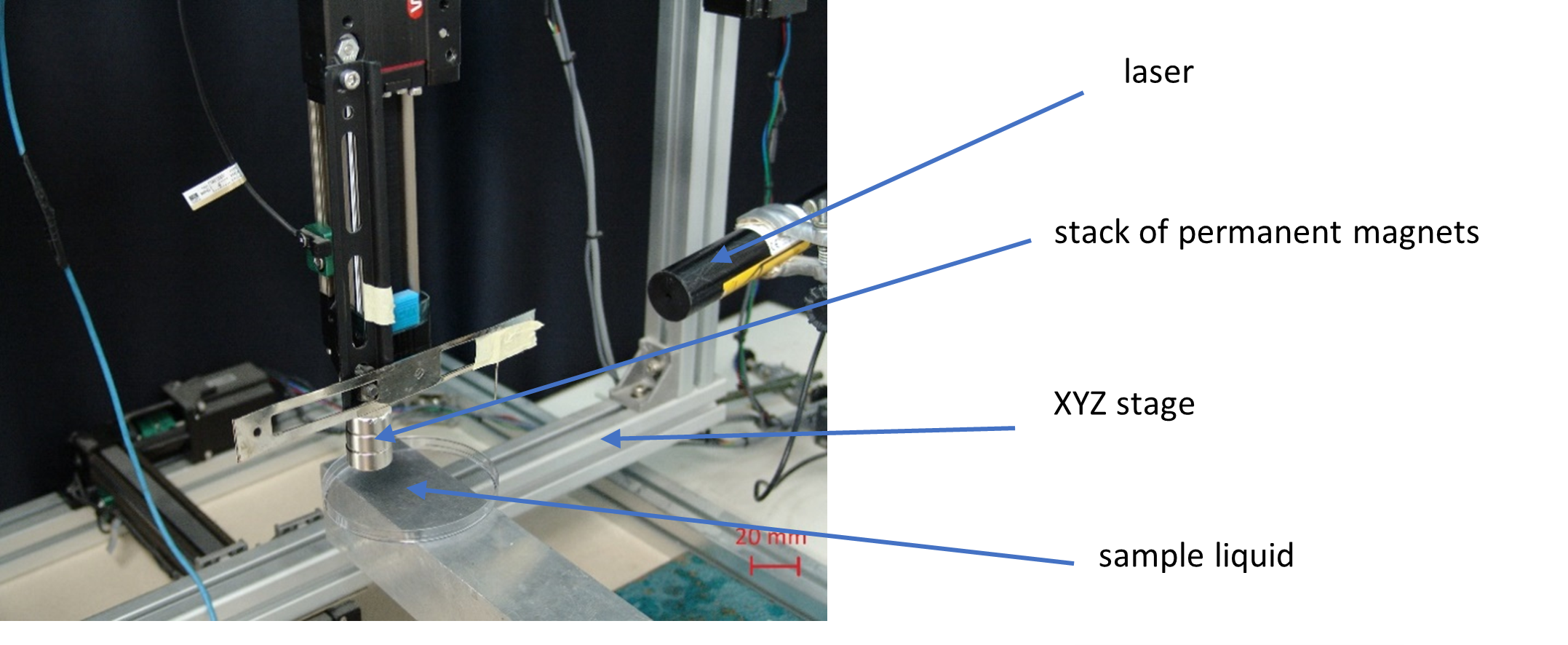}
\caption{The set-up of the experimental unit.}
\label{fig:4}

\includegraphics[scale=1]{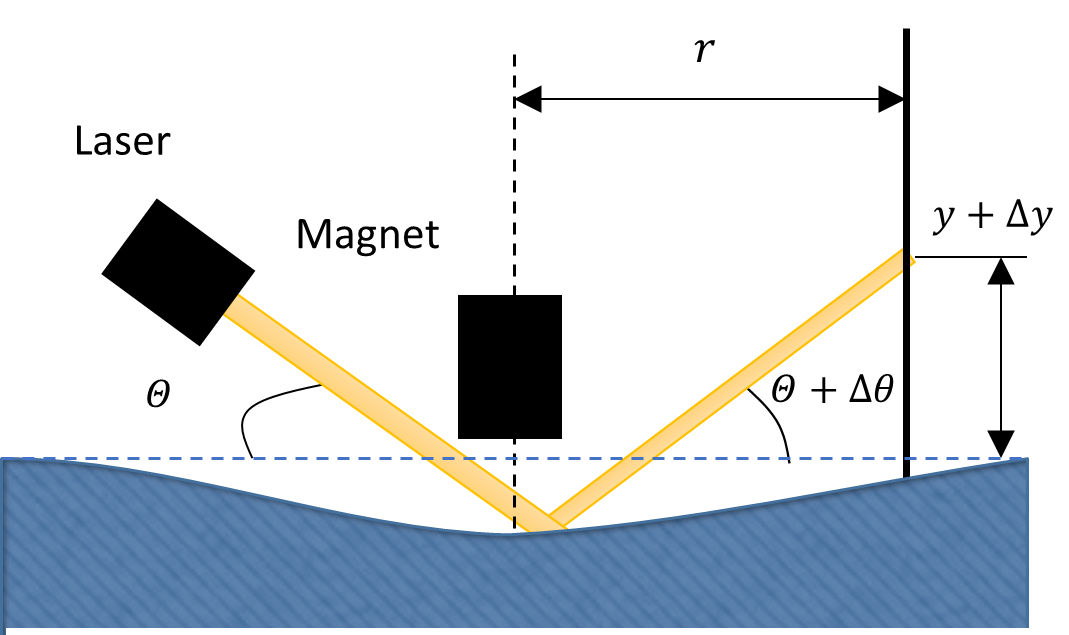}
\caption{Scheme of the experimental unit used for the measurement of the magnetic susceptibility of liquids.}
\label{fig:5}

\includegraphics[scale=1]{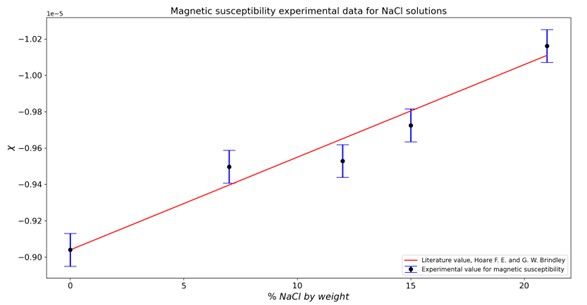}
\end{figure*}
\begin{figure*}
\includegraphics[scale=1]{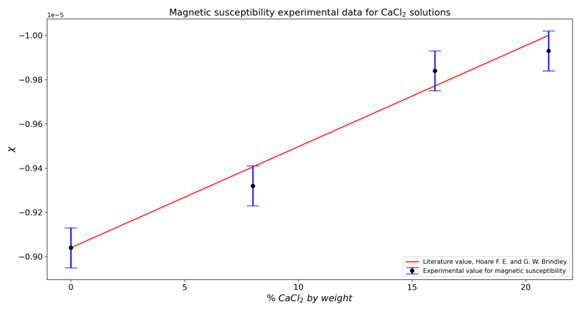}
\caption{Values of the magnetic susceptibility extracted from the experimental data (black circles) for various salt solutions in water are depicted. Red solid straight line depicts the literature value of the magnetic susceptibility taken from ref.[24].}
\label{fig:6}

\includegraphics[scale=1]{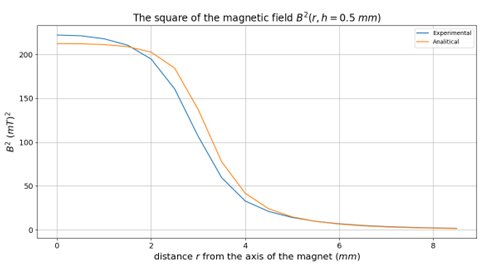}
\caption{The comparison of the squared magnetic field $B^2\left(r\right)=B_r^2(r)+B_z^2(r)$ as established experimentally (blue curve) and calculated (brown curve) with Eq. A1a-A1b at a distance of $h=0.5\ mm$ from the surface of the magnet is shown. The diameter of the magnet is 6 mm.}
\label{fig:7}
\includegraphics[scale=1]{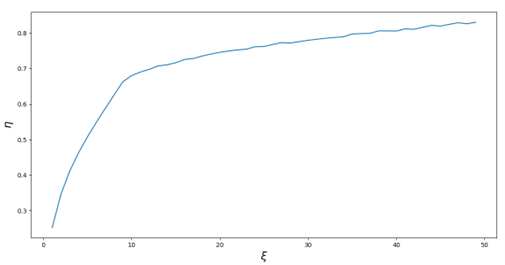}
\caption{The dependence of the parameter $\eta=\frac{\chi_1}{\chi_2}$ on the dimensionless number $\xi$.}
\label{fig:8}
\end{figure*}

\end{document}